\documentstyle{EuroPhys}

\def\etal{{\hbox{{\tenit\ et al.\/}\tenrm :\ }}}

\def\And{{\rm and\ }}

\newif\ifboo \boofalse

\def\Review#1{\boofalse{\it #1},}
\def\Name#1{{\sc #1},}
\def\Vol#1{\ifboo Vol. {\bf #1}\else{\bf #1}\fi}
\def\Year#1{\ifboo #1\else(#1)\fi}
\def\Book#1{\bootrue{\it #1},}
\def\Page#1{\ifboo {\rm p. #1}\else{\rm #1}\fi}

\input epsf
\newcommand{\be}{\begin{equation}}
\newcommand{\ee}{\end{equation}}
\begin{document}
\shorttitle{Amici \etal , Scaling properties of discrete fractals}
\title{Scaling properties of discrete fractals }
\author{A.Amici\inst{1}\footnote{E-mail address:amici@mpipks-dresden.mpg.de} 
\And M. Montuori\inst{2}\footnote{E-mail address:marco@pil.phys.uniroma1.it }}
\institute{
    \inst{1} Max-Planck-Institut f\"{u}r Physic komplexer Systeme,
     01187 Dresden, Germany \\
 \inst{2}INFM Sezione Roma I, Dipartimento di Fisica, Universit\`a di Roma
``La Sapienza'', 00185 Roma, Italy \\}
\rec{ }{ }
\pacs{
\Pacs{64}{60}{Fractals}
\Pacs{5}{20}{Statistical Mechanics}
\Pacs{5}{40}{Fluctuation Phenomena}
}
\maketitle

\begin{abstract}
An important problem in the analysis of experimental data showing fractal 
properties, is that such samples are composed by a set of points limited 
by an upper and a lower cut off.
We study how finite size effect due to the discreteness of the 
sets may influence self similar properties even far from these cut-offs.
Estimations of these effects are provided on the basis of the
characteristics of the samples. In particular we present an estimate of the 
length scale above which information about average quantities 
is reliable, by explicitly computing 
discreteness effects in number counting .
The results have particular importance
in the statistical analysis of the distribution of galaxies.

\end{abstract} 
\vspace{0.5cm}

The simplest way to characterise a fractal structure 
is to analyse the scaling behaviour 
on large sets of data, of convenient 
averaged quantities, such
as the two point correlation function 
\cite{man1,falco}. 
Here, we introduce 
a more complete description of self similar structures
by including the properties of the fluctuations of
such quantities with respect to the average values. 
Previous approaches to this problem include 
the analysis of the void distribution 
\cite{manlac,GEF} and analysis of
the tree points correlation function \cite{BLU}.

In order to introduce a theory of the 
fluctuations in fractal patterns, it is crucial to 
characterize the finite size effects \cite{gras}.
Real data \cite{avnir} are 
sets of a finite number of points where
fluctuation analysis requires much larger samples than 
those needed to compute the averaged quantities.
In particular in the field of large scale structure
astrophysics, due to observational limitations,
unaveraged quantities are widely used in the
statistical analysis of galaxy correlations
\cite{man1,peebles,syl97}.

In this work we further develop 
fluctuation analysis and give quantitative criteria to 
define its statistical significance for a given data set. 
In order to 
proceed with this analysis we distinguish between finite 
size effects
related to the shape of the data set boundaries and the 
finite size effects
due to the discreteness of the sample. In order to avoid 
the former we
carefully select the region for number counting.
 To approach the latter we compute the probability 
distribution of fluctuations in number
counts on a discrete set of points.

 The scaling quantity we will consider 
is the conditional average number mass $N(r)$ 
defined below. Given a set of $N_{tot}$ points,
let $n_i(r)$
be the Conditional number Mass (CM) from the point $i$
defined as follows:
\begin{equation}
\label{eq1}
n_i(r)=\int_{S(r,i)} \sum_j \delta({\bf x - x}_j)
d^d{\bf x}\; ,
\end{equation}
where $j$ runs over all the points in the set and
$S(r,i)$ is the $d$-dimensional hypersphere with radius $r$ 
centred at the point $i$.
$N(r)$ at a given $r$ is 
defined as the average of the CM's over
all the points whose distance 
from the sample boundaries $\bar{r}_i$
is greater than $r$:
\begin{equation}
\label{eq2}
N(r) = \left<n_i(r)\right>_{\bar{r}_i>r}\; .
\end{equation}
The supplementary condition is needed in order to avoid the 
finite size effects linked to the geometrical shape of the sample as 
pointed out in Ref.s \cite{col92,syl97}. 
With this restriction we now can face the problem of the discreteness of the 
data set.

\begin{figure}[h]
\epsfxsize=5cm
\vbox to 4cm{\vfill\centerline{\fbox{fig1.ps}}\vfill}
\epsffile{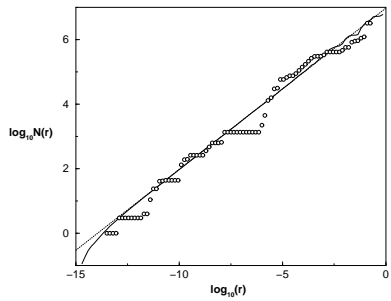}
\caption{ The solid line is the logarithmic conditional average number of
points as function of the logarithmic distance for a random $\beta$ model
with dimension $D=2$ and $10^7$ points.
Open circles are the counts from one point picked randomly.
The dotted line is the theoretical power law expected. 
The small deviations from the power law at small and large scales
are due to different kinds of finite size effects.
}
\label{fig1}
\end{figure}

For a fractal set of dimension $D$, $N(r)$ takes the form \cite{man1}:
\begin{equation}
\label{eq3}
N(r) = Br^{D}\; ,
\end{equation}
where $B$ is a prefactor independent from $D$.
In fig. \ref{fig1}, we compare 
the behaviour of a single CM to $N(r)$
for one of the fractal sets described below.
The multiplicative character of fluctuations
is reflected in their constant amplitude in
the log-log plot.
In order to compare fluctuations
at different scales we introduce the normalized
conditional number:
\begin{equation}
\label{eq4}
f_i(r) = \frac{n_i(r)}{N(r)}\; .
\end{equation}
For $N(r)\gg1$ it is possible to introduce a continuous representation
for the variable $f$, whose associated probability distribution
${\cal P}(f)$ is independent from $r$ due to the self-similarity.
On the other hand one expects the continuous limit 
does not to hold at a scale 
close to the lower cut-off. 
This is because in such a range the CM can take only a small set of discrete 
values corresponding to the presence of $0,1,2...$ points inside the
$d$-dimensional hypersphere. For this reason, it is possible to
observe fluctuations not scale invariant at small scales even if
the averaged quantities show the correct scaling behaviour.
The effect of the discreteness on fluctuations is particularly
evident when fractal correlations are present at distances
much smaller than the typical distance between neighbor points.
Samples with this property are equivalent to sets of
randomly picked points from much larger fractals. 

Generally,
the probability $\Pi(n,r)$ of find $n$ points inside a $d$-dimensional 
hypersphere of radius $r$ centered on a point of the set,
is the convolution of the scale invariant distribution density ${\cal P}(f)$
with a Poisson distribution $P_{m}(n)$ with average $m= fN(r) =fBr^D$:
\begin{equation}
\label{eq5}
\Pi(n,r) = \int \limits_0^{\infty} df {\cal P}(f) P_{fN(r)}(n) =
\left< P_{fN(r)}(n) \right>_f\; ,
\end{equation}
where the notation $\left< \phi(f) \right>_f \equiv
\int \limits_0^{\infty} df {\cal P}(f)\phi(f)$ 
stands for the average with respect
to the distribution function ${\cal P}(f)$.
We define also $\left< \phi(n) \right>_n \equiv \sum_{n=0}^{\infty}
P_{fN(r)}(n)\phi(n)$ for the average with respect to the Poisson
distribution.
The variance of the number of points $n$ is then:
\begin{equation}
\label{eq6}
\sigma^2(r)=\sum_{n=0}^{\infty}\Pi(n,r)n^2 - N^2(r)=
\left< \left< n^2 \right>_n \right>_f - N^2(r)\; ,
\end{equation}
where in the last equality we exchanged the sum with the integral.
Inserting eq.
(\ref{eq5})
in eq. (\ref{eq6}), 
adding and subtracting the quantity 
$\left<\left(fN(r)\right)^2 \right>_f$ we obtain:
\begin{equation}
\label{eq7}
\sigma^2(r)
=\left< \left< n^2 - f^2N^2(r)
\right>_n \right>_f + \left<
f^2N^2(r) - N^2(r) \right>_f\; .
\end{equation}
This gives:
\begin{equation}
\label{eq8}
\sigma^2(r)= N(r) + N^2(r)\sigma^2_{si}\; ,
\end{equation}
where $\sigma^2_{si}$ is the variance of scale invariant
distribution ${\cal P}(f)$. 
In the first term of eq. \ref{eq7}, the inner 
average is the variance of poisson 
distribution, i.e. $fN(r)$, and 
the outer average with respect 
$f$ will give $N(r)$.

This expression gives the 
relation between the variance of the number of points
in a discrete fractal set with respect the 
intrinsic variance of the fractal probability 
distribution ${\cal P}(f)$.
The variance of the fluctuations is then:
\begin{equation}
\label{eq9}
\sigma^2_f(r)\equiv\frac{\sigma^2(r)}{N^2(r)}=\sigma^2_{si} + \frac{1}{N(r)}\; ,
\end{equation} 
where the effect of the random sampling is simply to add the
Poisson contribution $1/N(r)$ to the scale invariant term
$\sigma^2_{si}$.
Through the expression in eq.(\ref{eq9}) it is possible to evaluate 
$\sigma^2_{si}$ at any scale. 
Moreover we can estimate the scale beyond which 
the poisson fluctuations are negligibile 
with respect the intrinsic ones.
In order to have the Poisson
term i.e. $10$ times smaller than the intrinsic one,
the CM should be fitted for $r$ larger than the minimal
statistical length $\lambda$:
\begin{equation}
\lambda=\left( \frac{10}{B\sigma_{si}^2} \right)^{\frac{1}{D}}\; .
\label{eq10}
\end{equation}
Eq. (\ref{eq10}) with $\sigma_{si}^2=1$ 
has been derived in a phenomenological way
in the context of large scale structure of the Universe
\cite{mont,syl97}.
The quantity $\sigma^2_f(r)$ has been called
lacunarity \cite{falco}, and is usually referred to as one of the
intrinsic characterisation of a fractal set. However the scale invariant
value $\sigma_{si}^2$ appear only in the $r \gg \lambda$ region.

Having defined this quantity, it is possible to further characterise the
fractal set. This is done by noticing that fluctuations show a 
characteristic ``correlation length'' on a $log-log$ scale. This corresponds
to a correlation scale difference $\Delta S$ above which
fluctuations are uncorrelated
and can be used as an estimate of limiting
``length'' that has to be considered in order to deal with reliable data.

The correlation function between fluctuations at
scales $r_1$ and $r_2$ is:
\begin{equation}
\label{eq11}
\tilde{S}(r_2,r_1) 
=\left< (f(r_2)-1)(f(r_1)-1)  \right>
\equiv \left< \delta f(r_2) \delta f(r_1)  \right>\; .
\end{equation}
In the self similar regime one expects $\tilde{S}$ 
to be a function of the ratio of the scales, i.e. 
$\tilde{S}(r_2,r_1)=S(\log_{10}(r_2/r_1))\equiv S(\Delta)$.
An estimation for $S(\Delta)$  may be found in the following way:
let $n_i(r_1)$ the CM from the point $i$ up 
to a scale $r_1$, $n_i(r_2)$ the CM up to a 
scale $r_2>r_1$ and $\tilde {n_i(r_1,r_2)}$
the number of points in the shell $r_2-r_1$.
We can express $\delta f_i(r_2)$ in the following way:
\begin{eqnarray}
\label{eq12}
\delta f_i(r_2) & = & f_i(r_2) - 1 = \frac{n_i(r_2)}{N(r_2)} -1
	 =\frac{n_i(r_1) +\tilde n(r_1,r_2)}{N(r_1)}\cdot \frac{N(r_1)}{N(r_2)} -1 =\nonumber  \\
       	&=&{f_i(r_1)}\frac{r_1^D}{r_2^D}-1 +\frac{\tilde n(r_1,r_2)}{N(r_2)} = 
	\delta f_i(r_1) + \delta \tilde f(r_1,r_2)\; . \nonumber \\
\end{eqnarray}
Assuming that
the number of points in the $r_2>r_1$ shell, 
$\tilde n_i(r_1,r_2)$ is uncorrelated with $n_i(r_1)$, we 
have that  $\left< \delta f_i(r_1) \delta \tilde f_i(r_1,r_2) \right>=0$.
Therefore:
\begin{equation}
\label{eq13}
S\left(\Delta\right)\! 
= \! \left< \!\!\delta f(r_1) \left( \delta f(r_1)\frac{r_1^D}{r_2^D}
\right)\!\!  \right> 
= \sigma^2_f \left( \frac{r_1}{r_2} \right)^D\!\!
= \sigma^2_f \cdot 10^{-\frac{\Delta}{\Delta_c}}\; ,
\end{equation}
where $\Delta_c=D^{-1}$.
In the general case $\left< \delta f_i(r_1)\delta\tilde{f_i}\right> >0$
and then eq. (\ref{eq13}) is the lower limit.

To show the applicability of the formulae derived 
we will analyse selected fractal samples.
Those fractal sets are obtained picking 
randomly points from larger samples generated by a random $\beta$ model (RBM)
algorithm \cite{vulp} in two dimensions. 
The original sets contain from $10^9$ to $10^{10}$ points, 
while the diluted samples contain around $10^7$ points, in 
a square domain of side 2. 

Two characteristic lengths for the sets are respectively the minimal 
possible distance between points $\epsilon$ (which corresponds also to the 
minimal distance at which one observes fractal correlations), and
the average distance to the nearest point $r_0$, which is related
to the prefactor $B$ through $r_0^D\sim B^{-1}$. Whilst the two lengths 
may differ substantially for highly diluted samples, they are almost identical 
for the RBM without dilution.

In fig \ref{fig1}, we show $N(r)$ for a RBM fractal 
with $D=0.5$, $\epsilon\sim10^{-15}$, $r_0\sim 10^{-14}$.
The dotted line is the theoretical 
behaviour for $N(r)$ in this sample and the open circles are
the values for the CM from a randomly chosen point. 

\begin{figure}[h]
\epsfxsize=5cm
\vbox to 4cm{\vfill\centerline{\fbox{fig2.ps}}\vfill}
\epsffile{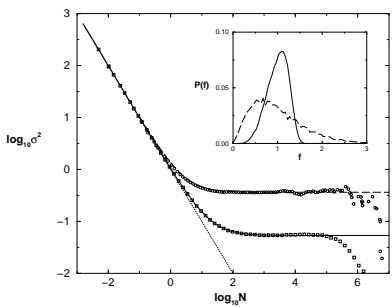}
\caption{Reduced variance vs. the average number of points,
logarithmic scale. Squares refer to a RBM with $D=1.8$
and circles to one with $D=0.5$.
The solid and the dashed lines are fitted
to eq. (\protect\ref{eq9}). In the inset are shown the numerical
${\cal P}(f)$ for the two sets above.}
\label{fig2}
\end{figure}

In fig. \ref{fig2} is reported the quantity $\sigma_f^2(r)$ 
for two sets of data: 
the fractal with $D=0.5$ and another one with $D=1.8$
($\epsilon\sim10^{-5}$ and $r_0\sim10^{-4}$).
In eq.(\ref{eq9}) we take as independent 
variable the $N(r)$; in this way it is possible to 
fit the data to eq. (\ref{eq9}) with a single parameter $\sigma^2_{si}$
(solid lines). The $N^{-1}(r)$ scaling is the Poissonian 
term and the plateau gives the intrinsic value $\sigma^2_{si}$.
At large $N(r)$ deviations from the theoretical behaviour
are observed. The reason for that may be understood through
the following argument. All the points belonging to a cluster
of size $r_c$ have correlated $n_i(r)$ for $r > r_c$.
Sample independent behaviour of fluctuations is obtained
for $r\sim r_c$ as long 
as a large number of independent clusters of size
$r_c$ is contained in the sample volume. 
This condition is less and less fulfilled as $r$ increases. 
This gives rise to a strong reduction of $\sigma^2_f(r)$
whilst approaching the sample size.
In the insert, we show the distributions
of the fluctuations in the scale invariant regions, ${\cal P}(f)$,
for the two samples. 

\begin{figure}[h]
\epsfxsize=5cm
\vbox to 4cm{\vfill\centerline{\fbox{fig3.ps}}\vfill}
\epsffile{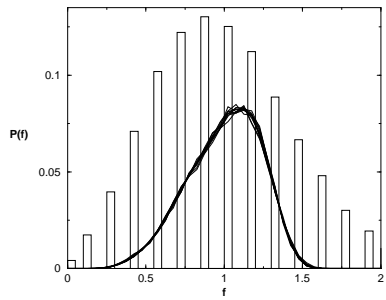}
\caption{Solid lines refer to the numerical probability density of
the fluctuation at several different scales within the range
of self similarity ($D=1.8$). The histogram is the same quantity
at smaller scales, in the regime where the discretization of $f$
and the Poisson noise are much more relevant.}
\label{fig3}
\end{figure}

In fig. \ref{fig3} we show the 
probability distribution for the fluctuations $f$ at
various $r$, for the $D = 1.8$ fractal. Continuous lines 
refer to $r$
in the range $[10^{-3}$-$10^{-2}]$ where the good superposition
of the different curves signals the scale invariance of the fluctuations.
Outside from this region the measured distribution deviates from
the intrinsic shape.
The histogram of fig \ref{fig3} is 
the probability distribution at smaller scales ($\sim10^{-4}$), 
out from the {\it scale invariant} regime .
The main features of it are the discreteness of the
allowed value for $f$ and the larger variance due to the Poisson
noise as explained before.

\begin{figure}[h]
\epsfxsize=5cm
\vbox to 4cm{\vfill\centerline{\fbox{fig4.ps}}\vfill}
\epsffile{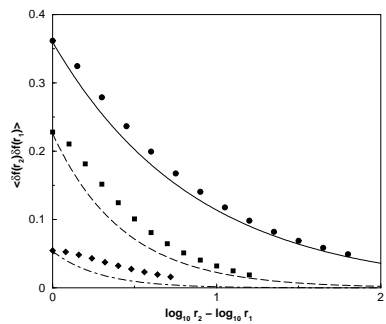}
\caption{Correlation between fluctuation versus the scale difference.
Circles, squares and diamonds refer to fractals 
with $D=0.5,1,1.8$ respectively. Lines are fitted to eq. (\protect\ref{eq13}).}
\label{fig4}
\end{figure}

Fig \ref{fig4} shows the measure of correlation 
between fluctuations $S(r_1/r_2)$ at different scales.
The points refer to three fractal sets with respectively
fractal dimension $D=0.5,1,1.8$ and are the average of 
$S(r_2/r_1)$ for all
$r_1$ and $r_2$ inside the scale invariant range.
While for small fractal dimension, $D=0.5$,
the values estimated from eq. (\ref{eq13})
are in rather good agreement with the data, the sets with
$D$ closer to the dimension of the embedding space
($D=1.8$) show higher additional correlations between fluctuations
at different scales. According to eq. (\ref{eq13})
we may estimate the correlation scale difference to be
$\Delta_c \sim D^{-1}$ for small fractal dimension.
On the other hand for large dimensions, $\Delta_c$
appears not to decrease below $1$. An estimation for 
the correlation scale difference
in the two dimensional RBM is:
\begin{equation}
\label{eq14}
\Delta_c = \mbox{Max} \left(1,\,\frac{1}{D} \right)\; .
\end{equation}
Recovering the average behaviour 
of $N(r)$ from the measure of single 
$n_i(r)$ requires then a range of scale 
larger then $\Delta_c$. 

The main result is that this analysis allows one to
specify the value of $\Delta_c$. Thereby introduces 
a measure for a quantity 
which has been customary 
to define as ``large enough'' \cite{avnir}.
For instance, the fractal 
with dimension $D= 0.5$ has $\Delta_c = 2$ and 
in fig. \ref{fig1} the corresponding CM shows 
typical length of fluctuations of $2-3$ 
orders of magnitude. 

A direct application of such results is 
in the study of statistical properties 
of galaxy distribution.
Here, a fundamental measure is the 
the counts of galaxies versus the distance 
from the Earth, or galaxy CM, $n_E(r)$.
The interpretation of 
the behaviour of galaxy CM is 
crucial in the debate about the 
existence or not of the homogeneity scale 
of galaxy distribution in the Universe, 
which is one of the pillars of standard Big Bang theory.
On the other side, the amount of 
available data are not so large, making 
important to check for finite size effects.

In conclusion, the intrinsic distribution 
of the fluctuations ${\cal P}(f)$ can be measured only for 
scales greater than the minimal statistical 
length, defined in eq. (\ref{eq10}).
On the other hand, the lacunarity of a discrete fractal may be
evaluated, at all scales, as $\sigma_f^2(r)$.
In those cases in which the CM has to be
used in extracting fractal parameters from experiments,
data for $r<\lambda$ have to be discarded and
a range of scale larger than $\Delta_c$ is needed.

It is a great pleasure to acknowledge 
G.Caldarelli for critical reading of 
the manuscrpt and for many 
suggestions and discussions. We 
would also acknowledge A. Gabrielli, M.Joyce, 
M. Mu\~noz, L. Pietronero 
and F.Sylos Labini for valuable discussions and
critical comments.
 This work has been partially supported by 
EEC TMR Network ``Fractal Structure and 
self-organization'' ERBFMRXCT980183.

\end{document}